\documentstyle[aps,prl,array,epsf]{revtex}
\begin{document}
\def\beq{\begin{equation}}
\def\eeq{\end{equation}}
\def\bea{\begin{eqnarray}}
\def\eea{\end{eqnarray}}
\newcommand{\ket}[1]{| #1 \rangle}
\newcommand{\bra}[1]{\langle #1 |}
\newcommand{\braket}[2]{\langle #1 | #2 \rangle}
\newcommand{\tr}{\rm tr}
\newcommand{\rank}{\rm rank}
\newcommand{\proj}[1]{| #1\rangle\!\langle #1 |}
\newcommand{\ba}{\begin{array}}
\newcommand{\ea}{\end{array}}
\newtheorem{theo}{Theorem}
\newtheorem{defi}{Definition}
\newtheorem{lem}{Lemma}
\newtheorem{exam}{Example}
\newtheorem{prop}{Property}
\newtheorem{coro}{Corollary}

\twocolumn[\hsize\textwidth\columnwidth\hsize\csname
@twocolumnfalse\endcsname
\title{Entanglement-Assisted Classical Capacity of Noisy Quantum Channels}
\author{Charles H. Bennett$^1$,
Peter W. Shor$^2$, John A. Smolin$^1$, and
Ashish V. Thapliyal$^3$}
\address{$^1$IBM Research Division, Yorktown Heights, NY 10598, USA 
{\tt bennetc, smolin@watson.ibm.com} $^2$AT\&T Research, Florham
Park NJ 07932 {\tt shor@research.att.com}
$^3$Physics Dept., Univ. of California, Santa Barbara, CA 93106,
USA {\tt ash@physics.ucsb.edu}}

\date{\today}

\maketitle
\begin{abstract}

Prior entanglement between sender and receiver, which exactly doubles
the classical capacity of a noiseless quantum channel, can increase the
classical capacity of some noisy quantum channels by an arbitrarily large
constant factor depending on the channel, relative to the best known
classical capacity achievable without entanglement. The enhancement
factor is greatest for very noisy channels, with positive classical
capacity but zero quantum capacity.  We obtain exact expressions for the
entanglement-assisted capacity of depolarizing and erasure channels in
$d$ dimensions.

\end{abstract}

% close bracket for pretty quant-ph mode
]
%
%\narrowtext

Prominent among the goals of quantum information theory are
understanding entanglement and calculating the several capacities
of quantum channels. Physically, a quantum channel can be pictured
as the transfer of some quantum system from sender to receiver. If
the transfer is intact and undisturbed, the channel is noiseless;
if the quantum system interacts enroute with some other system, a
noisy quantum channel results. Quantum channels can be used to
carry classical information, and, if they are not too noisy, to
transmit intact quantum states and to share entanglement between
remote parties. Unlike classical channels, which are adequately
characterized by a single capacity, quantum channels have several
distinct capacities. These include a classical capacity $C$, for
transmitting classical information, a quantum capacity $Q$, for
transmitting intact quantum states, a classically-assisted quantum
capacity $Q_2$, for transmitting intact quantum states with the
help of a two-way classical side-channel, and finally $C_E$, the
{\em entanglement-assisted classical capacity,} which we define as
a quantum channel's capacity for transmitting classical information with 
the help of unlimited prior pure entanglement between sender and 
receiver~\cite{bpv98}. In most cases, only upper and lower bounds on 
these capacities are known, not the capacities themselves~\cite{bdsbst}. 

Entanglement, eg in the form of Einstein-Podolsky-Rosen (EPR)
pairs of particles shared between two parties, interacts in subtle
ways with other communications resources.  By itself, prior
entanglement between sender and receiver confers no ability to
transmit classical information, nor can it increase the capacity
of a classical channel above what it would have been without the
entanglement.  This follows from the fact that local manipulation of one
of two entangled subsystems cannot influence
the expectation of any local observable of the other
subsystem~\cite{bbcjpw93,note}. This is sometimes loosely called
the constraint of causality, because its violation would make it
possible to send messages into one's past.

On the other hand, it is well known that prior entanglement can
enhance the classical capacity of {\em quantum\/} channels. In the
effect known as superdense coding, discovered by
Wiesner~\cite{bw92}, the classical capacity of a noiseless quantum
channel is doubled by prior entanglement. In other words,
$C_E\!=\!2C\,$ for any noiseless quantum channel. We show that for
some channels this enhancement persists, and even increases, as
the channel is made more noisy, even after the channel has become
so noisy that its quantum capacities $Q$ and $Q_2$ both vanish,
and the channel itself can be simulated by local actions and
classical communication between sender and receiver~\cite{bpv98}.

This is perhaps surprising, since it might seem that any quantum
channel that can be classically simulated ought to behave like a
classical channel in all respects---in particular not having its
capacity increased by prior entanglement.  In fact there is no
contradiction, because, as we shall see, even when a quantum
channel can be classically simulated, the simulation necessarily
involves some amount of forward classical communication from the 
sender (henceforth ``Alice'') to the receiver (``Bob'');
and this information is never less than the channel's
entanglement-assisted capacity. Thus for any quantum channel,
$C\leq C_E\leq FCCC$, where $FCCC$ denotes the forward classical
communication cost, ie the forward classical capacity needed, in
conjunction with other resources, to simulate the quantum channel.

To illustrate these inequalities consider a specific example, the
2/3-depolarizing qubit channel, which transmits the input qubit
intact with probability 1/3 and replaces it by a random qubit with
probability 2/3. As is well known, this noisy quantum channel,
sometimes referred to as the classical limit of teleportation, can
be simulated classically by the following ``measure/re-prepare"
procedure: A third party chooses a random axis R and tells both
Alice and Bob. Then Alice measures the input qubit along this axis
and tells Bob the one-bit result, after which Bob prepares an
output qubit in the same state found by Alice's measurement.
Evidently the FCCC of this procedure is 1 bit, but the best known
classical capacity of a 2/3 depolarizing channel (realized by
encoding 0 and 1 as $\ket{0}$ and $\ket{1}$ on the input side and
measuring in the same basis on the output side) is about 0.0817
bits, the  capacity of a classical binary symmetric channel of
crossover probability 1/3.  As we shall show, the $C_E$ of the 2/3
depolarizing channel is about 0.2075 bits, more than twice the unassisted
value, but still safely less than the 1 bit forward
classical cost of simulating the channel by measure/re-prepare, which we
denote $FCCC_{MR}$.

Suppose we wished to simulate not a 2/3-depolarizing channel, but a
5/6-depolarizing channel. Clearly this could be done by simulating the
2/3-depolarizing channel then further depolarizing its output. But a more
economical simulation would be for Alice to send her one-bit measurement
result to Bob not through a noiseless classical channel but through a
noisy classical channel of correspondingly lesser capacity. If she sent
it through a binary symmetric channel of randomization probability 1/2
(equivalent to a crossover probability 1/4), the 5/6-depolarizing
channel would be have been simulated at an $FCCC_{MR}$ of only
$1\!-\!H_2(1/4)\approx 0.1887$ bits per channel use, where $H_2$ is the
binary Shannon entropy $H_2(p)\!=\!-p\log_2p\!-\!(1\!-\!p)\log_2(1\!-\!p)$.
This is of course greater than the 5/6-depolarizing channel's best known
classical capacity of $1\!-\!H_2(5/12)\approx 0.02013$. The 5/6-depolarizing
channel's entanglement-assisted capacity must lie between these two
bounds.

We now develop these ideas further to obtain an exact expression
for $C_E$ for an important class of channels, the\/
$d$-dimensional depolarizing channel ${\cal D}^{(d)}_x$ of
depolarization probability $x$. This is the channel that transmits
a $\,d$-state quantum system intact with probability $1\!-\!x$ and
randomizes its state with probability $x$. We show that in the
high-depolarization limit $x\!\rightarrow\!1$ this channel's
entanglement-assisted capacity is $d\!+\!1$ fold higher than the
best known lower bound on the classical capacity of the same
channel without prior entanglement. This lower bound, the
``one-shot'' classical capacity $C_1,$ is defined as the maximum
classical information that can be sent through a {\em single\/}
use of the channel, without prior entanglement, by an optimal
choice of source states at the channel input and an optimal
measurement at the channel output. For this highly symmetric
channel, this optimum can be achieved by assigning equal
probability $1/d$ to each state of an arbitrary orthonormal basis
$\{\ket{0},\ket{1}\ldots\ket{d\!-\!1}\}$ at the channel input, and
performing a complete von Neumann measurement in the same basis at
the channel output. This causes the quantum channel to behave as a
\/ $d$-ary symmetric classical channel of randomization
probability $x$, giving a capacity
\beq C_1({\cal D}^{(d)}_x) =
\log_2d -H_d(1\!-\!x\frac{(d\!-\!1)}{d}), \label{ddepc1}
\eeq
where $H_d(p) =
-p\log_2(p)-(1\!-\!p)\log_2((1\!-\!p)/(d\!-\!1))$ is the Shannon entropy of
a \/ $d$-ary distribution consisting of one element of probability
$p$ and $d\!-\!1$ elements each of probability
$(1\!-\!p)/(d\!-\!1)$. This input ensemble is known to be optimal,
for a one-shot use of the channel, because it saturates the Holevo
bound $C_1\leq \log_2d-\overline{S(\rho_i)}$, on the one-shot
capacity \cite{holevo}, where $\overline{S(\rho_i)}$ is the
average von Neumann entropy of the output states $\rho_i$.
\begin{figure}[tb]
\epsfxsize=6.7cm \epsfbox{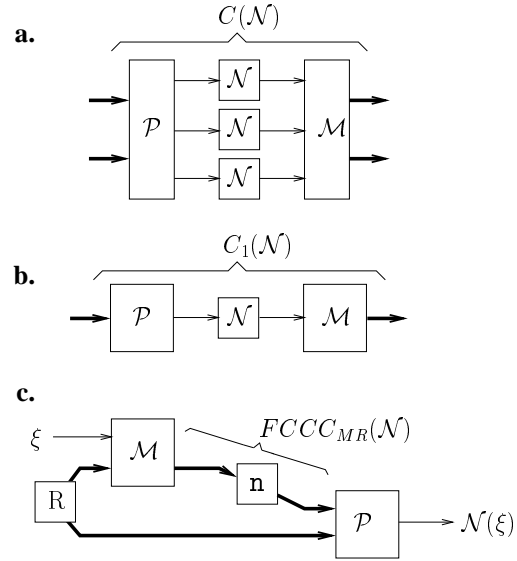}\medskip \caption{a) The
classical capacity $C$ of a quantum channel ${\cal N}$ is the
optimal asymptotic input:output mutual information per channel use
achievable by a preparator ${\cal P}$ mapping classical inputs
(thick lines) to a possibly entangled input state supplied to
multiple instances of the quantum channel, and a collective
measurement ${\cal M}$ mapping the possibly entangled output state
back to classical data. b) The one-shot classical capacity $C_1$
is the maximum classical mutual information between input
preparations and output measurement results for a single use of
the channel.  c) Some quantum channels ${\cal N}$, acting on an
unknown quantum input $\xi$, can be simulated classically by
having Alice measure the quantum input in a common random basis R,
send Bob the result through a noiseless or noisy classical channel
{\large \tt n}, after which he uses it to re-prepare an
approximation ${\cal N}(\xi)$ to the quantum input $\xi$ at the
channel output. The forward classical communication cost
($FCCC_{MR}$) of this simulation is the classical (Shannon)
capacity of the internal classical channel {\large \tt n}.
Allowing {\large \tt n} to be noisy makes ${\cal N}$ more noisy,
but reduces $FCCC_{MR}$.} \label{defs1}
\end{figure}

\noindent 

Similarly, it is easy to generalize the measure/re-prepare
construction to show that a \/ $d$-dimensional depolarizing
channel can be simulated classically whenever $x\geq d/(d+1)$, at
a cost \beq FCCC_{MR}({\cal D}^{(d)}_x)=
\log_2(d)-H_d\left(d-x(d-(1/d))\right). \label{ddepFCCC} \eeq The
simulation is performed by having Alice measure in a pre-agreed
random basis, send Bob the result through a \/ $d$-ary symmetric
noisy classical channel, after which he re-prepares an output
state in the same basis. Figure 1 compares the definitions of
asymptotic capacity $C$ and one-shot capacity $C_1$, and
illustrates the measure/re-prepare technique for simulating some
noisy quantum channels classically.

So far, we have only given lower and upper bounds on $C_E$, without
calculating $C_E$ itself. To do so we use modified versions of the
well-known superdense coding~\cite{bw92} and
teleportation~\cite{bbcjpw93} protocols to obtain tighter lower and
upper bounds, respectively, which in the case of depolarizing and
erasure channels coincide, thereby establishing $C_E$ exactly for these
channels. We treat the case where ${\cal N}$ is a generalized
depolarizing channel ${\cal D}^{(d)}_x$ first.

Clearly $C_E$ for any noisy channel ${\cal N}$ can be
lower-bounded by the entanglement-assisted capacity via a
particular protocol, namely superdense coding with the noisy
quantum channel $\cal{N}$ substituted for the usual noiseless
return path for Alice's half of a shared maximally entangled EPR
state $\Psi$. This version of superdense coding is illustrated in
Figure 2a, and we shall use $C_{Sd}({\cal N})$ to denote the
entanglement-assisted capacity of ${\cal N}$ via this protocol.
Conversely (Figure 2b), $C_E({\cal N})$ can be upper-bounded by
the forward classical communication cost of simulating ${\cal N}$,
not by measure/re-prepare, but by a version of teleportation in
which the requisite amount of noise is introduced by substituting
a noisy classical channel \/{\tt N}\/ for the usual noiseless
classical arm of the teleportation procedure (the classical
channel \/{\tt N}\/ operates on a $d^2$-letter classical alphabet,
in contrast to the\/ $d$-letter alphabet used by the channel
\/{\tt\large n}\/ in the measure/re-prepare simulation of Fig 1c).
This upper bound follows from the fact that even in the presence
of prior shared entanglement, the FCCC of simulating a quantum
channel cannot be less than its classical capacity; otherwise a
violation of causality would occur. Whenever a quantum channel
${\cal N}$ can be simulated by teleportation with a noisy
classical arm we use $FCCC_{Tp}({\cal N})$ to denote the forward
classical communication cost of doing so.

In the case of depolarizing channels the two bounds coincide,
because of the readily verified fact that superdense coding and
teleportation map each $x$-depolarizing \/ $d$-dimensional quantum
channel into an $x$-randomizing $d^2$-ary symmetric classical
channel and vice versa.  Thus for all depolarizing channels ${\cal
D}^{(d)}_x$, \beq
C_E\!=\!C_{Sd}\!=\!FCCC_{Tp}\!=\!2\log_2d-\!H_{d^2}(1\!-\!x\frac{d^2\!-\!1}{d^2}).\label{ddepce}
\eeq From equations \ref{ddepc1} and \ref{ddepce} it can be seen
that in the high-noise limit $x\rightarrow 1$, the enhancement
factor $C_E/C_1$ approaches $d\!+\!1$. Thus prior entanglement can
increase classical capacity by an arbitrarily large factor. 
For large $d$, $C_E/C_1\approx 2$ for most $x$, rising sharply near $x=1$.

We now turn to the quantum erasure channel\cite{gbp}, which is
unusual among noisy quantum channels in that its capacities $C$,
$Q$ and $Q_2$ are known exactly \cite{bdsbst}.  A quantum erasure
channel transmits its \/ $d$-dimensional input state intact with
probability $1\!-\!x$ and with probability $x$
replaces the input by a unique $(d\!+\!1)$'st state,
called an erasure symbol, orthogonal to all the input states.  If
the channels ${\cal N}$ and \/{\tt N}\/ in Figure 2 are taken to
be, respectively, a\/ $d$-dimensional quantum erasure channel and
a\/ $d^2$-dimensional classical erasure channel, the superdense
coding and teleportation bounds can again easily be shown to
coincide, providing an entanglement-assisted capacity $C_E = 2
(1\!-\!x) \log d,$ exactly twice the erasure channel's ordinary
classical capacity. 
\begin{figure}[tb]
\epsfxsize=7.6cm \epsfbox{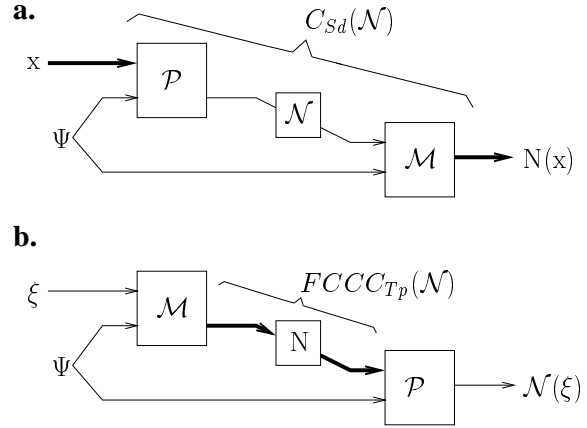}\medskip \caption{a) By using a
noisy quantum channel ${\cal N}$ in the protocol for superdense
coding, one obtains a lower bound $C_{Sd}$ on its
entanglement-assisted capacity $C_E({\cal N})$.  b) By using a
noisy classical channel \/{\tt N}\/ in the protocol for
teleportation to simulate a quantum channel ${\cal N}$, one
obtains an upper bound $FCCC_{Tp}$ on $C_E({\cal N})$. When these
two bounds coincide, they give $C_E({\cal N})$ exactly.}
\label{defs2}
\end{figure}
\noindent Figure 3 left shows all the capacities of the
quantum erasure channel. These capacities are of interest not only
in their own right, but also because they upper-bound the
corresponding capacities of the depolarizing channel, since a
quantum erasure channel can simulate a depolarizing channel by
having the receiver substitute a fully depolarized state for every
erasure symbol he receives. 

Returning to the depolarizing channel, we are in the peculiar
position of knowing its entanglement-assisted 
\vbox{\begin{figure} \epsfxsize=8cm
\epsfbox{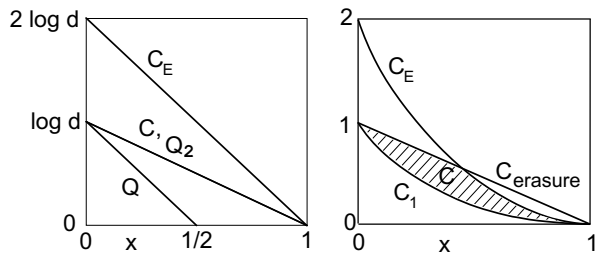}
\medskip
\caption{Left: Capacities of the quantum erasure channel.\\
Right: Bounds on the asymptotic classical capacity $C$ of the qubit depolarizing
channel.}
\end{figure}}
classical capacity $C_E$
without knowing its ordinary unassisted classical capacity $C$. The latter is
generally believed to be equal to the one-shot unassisted capacity
$C_1$, but the possibility cannot be excluded that a higher capacity
might be achieved asymptotically by supplying entangled inputs to
multiple instances of the channel (this cannot occur for $C_E$, where
any larger capacity would exceed $FCCC_{Tp}$, violating causality).  The
range of possible values for the depolarizing channel's unassisted
classical capacity $C$ is bounded below by its known
$C_1$, and above by its known $C_E$ and by the known unassisted
classical capacity $(1\!-\!x)\log d$ of the quantum erasure channel.  
Figure 3 right shows these bounds for the qubit case $d\!=\!2$. 

Although the depolarizing channel's unassisted capacity $C$
remains unknown in absolute terms for all $d$, the bounds
$C_1\!\leq\!C\!\leq\!(1\!-\!x)\log d$ become increasingly tight
{\em relative\/} to $C$ as $d\!\rightarrow\!\infty$, because, as
can readily be verified, the difference between the bounds
approaches $H_2(x)$ in this limit. Similarly, the depolarizing
channel's unassisted {\em quantum\/} capacity $Q$ is upper bounded
by the erasure channel's quantum capacity, $\max\{0,1\!-\!2x\}\log
d$, and lower bounded by the depolarizing channel's quantum
capacity via random hashing~\cite{bdsw}, $\log d\!-\!S([{\cal
N}\otimes I](\Psi))$. Here $[{\cal N}\otimes I](\Psi)$ is the
mixed state formed by sending half a maximally entangled $d\otimes
d$ pair $\Psi$ through the noisy channel. Again the difference
between the bounds approaches $H_2(x)$ as
$d\!\rightarrow\!\infty$.

The equality between $FCCC_{Tp}$ and $C_{Sd}$, which makes $C_E$
exactly calculable for depolarizing channels, holds for all
``Bell-diagonal'' channels~\cite{bdsw}, those that commute with
superdense coding and teleportation, so that $Tp(Sd({\cal
N}))={\cal N}$ \cite{note2}. For example the qubit dephasing channel, which
subjects its input to a $\sigma_z$ Pauli rotation with probability
$x/2$, has $C\!=\!1$ independent of $x$, while
$C_E=2\!-\!H_2(x/2)$. For other channels, it can be shown~\cite{long}
that
\beq 
C_E = \max_\Psi\{S(\rho)+S({\cal
N}(\rho))-S([{\cal N}\otimes I](\Psi))\}, \label{CEeq}
\eeq 
where $\Psi$ is a bipartite pure state in $d\otimes d$ and $\rho$ is 
its partial trace over the second party.  This capacity can be achieved 
asymptotically by applying superdense coding to a Schumacher-compressed 
version of $\rho^{\otimes n}$ for large $n$, and evaluating the resulting 
classical capacity by Holevo's formula~\cite{holevo}; that $C_E$ can
be no higher can be shown~\cite{long} using Holevo's formula and the
strong subadditivity property of quantum entropy.

A channel's entanglement-assisted {\em quantum\/} capacity $Q_E$
may be defined as its maximum rate for transmitting intact qubits
with the help of prior entanglement but no classical
communication. By teleportation and superdense coding, $Q_E=C_E/2$
for all channels.  Naturally, $Q_E$ upper bounds the unassisted
quantum capacity $Q$, but in most instances, eg the depolarizing
channel, tighter upper bounds are known.

We thank Howard Barnum, Herb Bernstein, David DiVincenzo, Richard
Jozsa, Barbara Terhal, Joy Thomas, and Bill Wootters for helpful discussions. CHB,
AVT, and JAS acknowledge support by the U.S. Army Research Office
under contract DAAG55-98-C-0041, and AVT under DAAG55-98-1-0366.


\begin{references}
\bibitem{note}As \protect\cite{bbcjpw93} showed, prior entanglement
cannot increase the capacity of a noiseless classical channel. To extend
this to the noisy case, we use the fact that (conversely to Shannon's
theorem) a noisy classical channel of capacity $C$ can be asymptotically
simulated by $C$ bits of noiseless forward communication, together with
prior random information $R$ shared between sender and receiver. (In our
application, $R$ does not represent an additional resource, because it
can be generated from the prior entanglement). Since we could not find one
in the literature, we sketch a proof of the converse Shannon theorem
adequate for the symmetric noisy classical channels used in this paper.
For each $\ell$-symbol block, the sender and receiver use the random
information $R$ to agree on a fresh random set $S_R$ of
$\approx\!2^{\ell(C+\epsilon)}\;$ $\ell$-symbol output words (eg a
random linear affine code). The sender then takes the $\ell$-symbol
input word $x$, and chooses an output word $y\!\in\!S_R$ with
probability $P'_R(y|x)=P(y|x)/(\sum_{y\in S_R}\!P(y|x))$ equal to the
noisy channel's native transition probability $P(y|x)$ renormalized over
the members of $S_R$. Using the pre-agreed code $S_R$, the output word
$y$ is transmitted to the receiver at a cost of $\ell(C+\epsilon)$ bits
of noiseless forward communication. Standard techniques can then be used
to show that for each $\epsilon\!>\!0$, the output distributions
$P'_R$ and $P$ converge as $\ell\!\rightarrow\!\infty$, and to extend
the proof to cover all discrete memoryless channels \protect\cite{long}.
\bibitem{note2} Note also that for Bell-diagonal quantum channels ${\cal N}$, 
the same classical channel ${\tt N}$ results whether the noise is applied in
the return path as in Figure 2a, or in either of the paths used to share the
entangled state $\Psi$.
\end{references}
\end{document}